\documentclass[conference]{IEEEtran}
\IEEEoverridecommandlockouts
\usepackage{epsfig,epstopdf,setspace,subfigure}
\usepackage{amsmath,amssymb,cite}
\newtheorem{proposition}{Proposition}

\newtheorem{theorem}{Theorem}

\begin{document}
\title{Coordination and Bargaining over the Gaussian Interference Channel}
\author{\IEEEauthorblockN{Xi Liu and Elza Erkip}
\IEEEauthorblockA{ECE Department, Polytechnic Institute of NYU, Brooklyn, NY 11201\\
Email: xliu02@students.poly.edu, elza@poly.edu}
\thanks{This material is based upon work partially supported by NSF Grant No.
0635177, by the Center for Advanced Technology in Telecommunications
(CATT) of Polytechnic Institute of NYU.}
}
\maketitle
\begin{abstract}
This work considers coordination and bargaining between two selfish users over a Gaussian interference channel using game theory. The usual information theoretic approach assumes full cooperation among users for codebook and rate selection. In the scenario investigated here, each selfish user is willing to coordinate its actions only when an incentive exists and benefits of cooperation are fairly allocated. To improve communication rates, the two users are allowed to negotiate for the use of a simple Han-Kobayashi type scheme with fixed power split and conditions for which users have incentives to cooperate are identified. The Nash bargaining solution (NBS) is used as a tool to get fair information rates. The operating point is obtained as a result of an optimization problem and compared with a TDM-based one in the literature.
\end{abstract}

\section{Introduction}

Interference channel (IC) is a fundamental model in information theory for studying interference in communication systems. In this model, multiple senders transmit independent messages to their corresponding receivers via a common channel. The capacity region or the sum-rate capacity for the two-user Gaussian IC is only known in special cases such as the strong interference case \cite{refereces:Sato81}\cite{references:Han81} or the noisy interference case\cite{references:Shang09}; the characterization of the capacity region for the general case remains an open problem. Recently, it has been shown in \cite{references:Etkin08} that a simplified version of a scheme due to Han and Kobayashi \cite{references:Han81} results in an achievable rate region that is within one bit of the capacity region of the complex Gaussian IC. However, any type of Han-Kobayashi (H-K) scheme requires full cooperation\footnote{Throughout the paper, ``cooperation'' means cooperation for the choice of transmission strategy including codebook and rate selection, which is different from cooperation in information transmission as in cooperative communications.} between the two users through the choice of transmission strategy. In practice, users are selfish in the sense that they choose a transmission strategy to maximize their own rates. They may not have an incentive to comply with a certain rule as in the H-K scheme and therefore not all rate pairs in an achievable rate region are actually attainable. When there is no coordination among the users, interference is usually treated as noise, which is information theoretically suboptimal in most cases.

In this paper, we study a scenario where two users operating over a Gaussian IC are selfish but willing to coordinate and bargain to get good and fair information rates. When users have conflicting interests, the problem of achieving efficiency and fairness could be formulated as a game-theoretical problem. The Gaussian IC was studied using noncooperative game theory in \cite{references:Etkin07}\cite{references:Larsson08}, where it was assumed that the receivers treat the interference as Gaussian noise. For the related Gaussian multiple-access channel (MAC), it was shown in \cite{references:Gajic08} that in a noncooperative rate game with two selfish users choosing their transmission rates independently, all points on the dominant face of the capacity region are pure strategy Nash Equilibria (NE). However, no single NE is superior to the others, making it impossible to single out one particular NE to operate at. The authors resorted to a mixed strategy which is inefficient in performance. Noncooperative information theoretical games were considered by Berry and Tse assuming that each user can select any encoding and decoding strategy to maximize its own rate and a Nash equilibrium region was characterized for a class of deterministic IC's \cite{references:Berry08}. Extensions were made to a symmetric Gaussian IC in \cite{references:Berry09}. Another game theoretical approach for interfering links is due to Han et al. \cite{references:Han05}, where the NBS from cooperative game theory was used as a tool to develop a fair resource allocation algorithm for uplink multi-cell OFDMA systems. Reference \cite{references:Leshem08} analyzed the NBS over the flat and frequency selective fading IC for time or frequency division multiplexing (TDM/FDM). The emphasis there was on the weak interference case. However, as we will show later, for the strong and mixed interference regimes, the NBS based on TDM/FDM may not perform very well, due to the suboptimality of TDM/FDM in those regimes.

In this work, assuming each user is selfish but willing to coordinate its action when an incentive exists, we formulate the interaction between the two users as a bargaining problem. We first illustrate how selfish users can bargain for a fair rate allocation over a Gaussian MAC. We then propose a two-phase mechanism for coordination between users over the Gaussian IC. First, the two users negotiate and only if certain incentive conditions are satisfied they agree to use a simple H-K type scheme with a fixed power split that gives the optimal or close to optimal sets of achievable rates\cite{references:Etkin08}. In the second phase, the NBS is used as a fairness criterion to obtain a preferred operating point over the achievable rate region. For all values of channel parameters, we study the incentive conditions for users to coordinate their transmissions. We also formulate the computation of the NBS over the H-K rate region as a convex optimization problem. Results show that the NBS exhibits significant rate improvements for both users compared with the uncoordinated case. The NBS obtained here can also achieve the maximum sum rate of the adopted H-K scheme in most cases, which demonstrates its strong efficiency.
\section{System Model}
\subsection{Channel Model}
In this paper, we focus on the two-user standard Gaussian IC
\begin{eqnarray}
Y_1 = X_1 + \sqrt{a}X_2+Z_1\\
Y_2 = \sqrt{b}X_1 + X_2 + Z_2
\end{eqnarray}
where $X_i$ and $Y_i$ represent the input and output of user $i \in \{1,2\}$, respectively, and $Z_1$ and $Z_2$ are i.i.d. Gaussian with zero mean and unit variance. Receiver $i$ is only interested in the message sent by transmitter $i$. Constants $\sqrt{a}$ and $\sqrt{b}$ represent the real-valued channel gains of the interfering links. If $a \geq 1$ and $b \geq 1$, the channel is {\em strong} Gaussian IC; if either $0<a<1$ and $b\geq 1$, or $0<b<1$ and $a\geq 1$, the channel is {\em mixed} Gaussian IC; if $0<a<1$ and $0<b<1$, the channel is {\em weak} Gaussian IC. We assume that transmitter of user $i$, $i \in \{1,2\}$, is subject to an average power constraint $P_i$. We let $\text{SNR}_i = P_i$ be the signal to noise ratio (SNR) of user $i$.  

\subsection{Achievable Rate Region}

The best known inner bound for the two-user Gaussian IC is the full H-K achievable region \cite{references:Han81}. Even when the input distributions in the H-K scheme are restricted to be Gaussian, computation of the full H-K region remains difficult due to numerous degrees of freedom involved in the problem \cite{references:Khandani09}. Therefore we assume users employ Gaussian codebooks with equal length codewords and consider a simplified H-K type scheme with fixed power split and no time-sharing as in \cite{references:Etkin08}. Let $\alpha \in [0,1]$ and $\beta \in [0,1]$ denote the fractions of power allocated to the private messages (messages only to be decoded at intended receivers) of user 1 and user 2 respectively. We define $\mathcal{F}$ as the collection of all rate pairs $(R_1,R_2)\in \mathbb{R}^2_{+}$ satisfying
\begin{equation}
R_1 \leq \phi_1 = C\left(\frac{P_1}{1+a\beta P_2}\right)
\end{equation}
\begin{equation}
R_2 \leq \phi_2 = C\left(\frac{P_2}{1+b\alpha P_1}\right)
\end{equation}
\begin{equation}
R_1 + R_2 \leq \phi_3 = \min\{\phi_{31},\phi_{32},\phi_{33}\}
\end{equation}
with
\begin{equation}
\phi_{31} = C\left(\frac{P_1+a(1-\beta)P_2}{1+a\beta P_2}\right) + C\left(\frac{\beta P_2}{1+b\alpha P_1}\right)
\end{equation}
\begin{equation}
\phi_{32} = C\left(\frac{\alpha P_1}{1+a\beta P_2}\right) + C\left(\frac{P_2+b(1-\alpha)P_1}{1+b\alpha P_1}\right)
\end{equation}
\begin{equation}
\phi_{33} = C\left(\frac{\alpha P_1+a(1-\beta)P_2}{1+a\beta P_2}\right) + C\left(\frac{\beta P_2+b(1-\alpha)P_1}{1+b\alpha P_1}\right)
\end{equation}
and
\begin{equation}
\begin{array}{l l}
2R_1+R_2 \leq \phi_4  = &\displaystyle C\left(\frac{P_1+a(1-\beta)P_2}{1+a\beta P_2}\right) + C\left(\frac{\alpha P_1}{1+a\beta P_2}\right)\\
&\displaystyle + C\left(\frac{\beta P_2+b(1-\alpha)P_1}{1+b\alpha P_1}\right)
\end{array}
\end{equation}
\begin{equation}
\begin{array}{l l}
R_1+2R_2 \leq \phi_5  = &\displaystyle C\left(\frac{P_2+b(1-\alpha)P_1}{1+b\alpha P_1}\right) + C\left(\frac{\beta P_2}{1+b\alpha P_1}\right)\\
&\displaystyle + C\left(\frac{\alpha P_1+a(1-\beta)P_2}{1+a\beta P_2}\right)
\end{array}
\end{equation}
where $C(x) = 1/2 \log_2(1+x)$.
The region $\mathcal{F}$ is a polytope and a function of $\alpha$ and $\beta$. We denote the H-K scheme that achieves the rate region $\mathcal{F}$ by $\text{HK}(\alpha,\beta)$. For convenience, we also represent $\mathcal{F}$ in a matrix form as $\mathcal{F} = \{\mathbf{R}|\mathbf{R} \geq \mathbf{0},\: \mathbf{R} \leq \mathbf{R}^1,\: \text{and}\:\mathbf{A}\mathbf{R} \leq \mathbf{B}\}$, where $\mathbf{R} = (R_1\:R_2)^t$, $\mathbf{R}^1 = (\phi_1\:\phi_2)^t$, $\mathbf{B} = (\phi_3\:\phi_4\:\phi_5)^t$, and
\begin{equation}
\mathbf{A} = \left(
\begin{array}{c c c}
1 & 2 & 1\\
 1 & 1 & 2
\end{array}
\right)^t
\end{equation}
Throughout the paper, for any two vectors $\mathbf{U}$ and $\mathbf{V}$, we denote $\mathbf{U} \geq \mathbf{V}$ if and only if $U_i \geq V_i$ for all $i$. $\mathbf{U} \leq \mathbf{V}$, $\mathbf{U} > \mathbf{V}$ and $\mathbf{U} <\mathbf{V}$ are defined similarly.

\subsection{Nash Bargaining Solution}
We employ the NBS as a criterion for selecting the desired operating point from a given achievable rate region, due to its Pareto optimality and fairness. 
In the following, we briefly review the basic concepts and results for the NBS in the context of our problem. More details are provided in Section III and Section IV.

We denote by $R^0_i$ the rate user $i$ would expect when both treat each other's signals as Gaussian noise. So we have $R^0_1 = C(\frac{P_1}{1+aP_2})$ and $R^0_2 = C(\frac{P_2}{1+bP_1})$. We choose $\mathbf{R}^0 = (R^0_1\:R^0_2)^t$ as the {\em disagreement point}, i.e., when negotiation breaks down, both users can transmit without cooperation at rates in $\mathbf{R}^0$. The bargaining problem can be represented by the pair $(\mathcal{F}, \mathbf{R}^0)$. We say $(\mathcal{F},\mathbf{R}^0)$ is {\em essential} iff there exists at least one allocation $\mathbf{R}'$ in $\mathcal{F}$ that is strictly better for both users than $\mathbf{R}^0$, i.e., the set $\mathcal{F} \cap \{\mathbf{R}|\mathbf{R}>\mathbf{R}^0\}$ is nonempty. In order for both users to have incentives for cooperation, it is required that $(\mathcal{F},\mathbf{R}^0)$ be essential; otherwise, at least one user does not have the incentive to bargain. A payoff allocation $\mathbf{R}$ is said to be \emph{Pareto optimal} iff there is no other allocation $\mathbf{R}'$ such that $R'_i \geq R_i, \forall i$, and $\exists i, R'_i > R_i$.



This bargaining problem is approached axiomatically by Nash \cite{references:Myerson91}. $\mathbf{R}^* = \mathbf{\Phi}(\mathcal{F},\mathbf{R}^0)$ is said to be an NBS in $\mathcal{F}$ for $\mathbf{R}^0$, if the following axioms are satisfied.
\begin{enumerate}
\item Individual Rationality: $\Phi_i(\mathcal{F},\textbf{R}^0) \geq R^0_i, \forall i$
\item Feasibility: $\mathbf{\Phi}(\mathcal{F},\mathbf{R}^0)\in \mathcal{F}$
\item Pareto Optimality: $\mathbf{\Phi}(\mathcal{F},\mathbf{R}^0)$ is Pareto optimal.
\item Independence of Irrelevant Alternatives: For any closed convex set $\mathcal{G}$, if $\mathcal{G} \subseteq \mathcal{F}$ and $\mathbf{\Phi}(\mathcal{F},\mathbf{R}^0) \in \mathcal{G}$,  then $\mathbf{\Phi}(\mathcal{G},\mathbf{R}^0) = \mathbf{\Phi}(\mathcal{F},\mathbf{R}^0)$.
\item Scale Invariance: For any numbers $\lambda_1, \lambda_2,\gamma_1$ and $\gamma_2$, such that $\lambda_1 > 0$ and $\lambda_2>0$, if $\mathcal{G} = \{(\lambda_1 R_1+ \gamma_1,\lambda_2 R_2 + \gamma_2)|(R_1,R_2)\in \mathcal{F}\}$ and $\mathbf{\omega} =(\lambda_1 R^0_1+ \gamma_1,\lambda_2 R^0_2 + \gamma_2) $, then $\mathbf{\Phi}(\mathcal{G},\mathbf{\omega}) = (\lambda_1 \Phi_1(\mathcal{F},\mathbf{R}^0)+ \gamma_1,\lambda_2 \Phi_2(\mathcal{F},\mathbf{R}^0) + \gamma_2)$.
\item Symmetry: If $R^0_1 = R^0_2$, and $\{(R_2,R_1)|(R_1,R_2)\in \mathcal{F}\} = \mathcal{F}$, then $\Phi_1(\mathcal{F},\mathbf{R}^0) = \Phi_2(\mathcal{F},\mathbf{R}^0)$.
\end{enumerate}

Axioms (4)-(6) are also called {\em axioms of fairness}.
{\theorem\cite{references:Myerson91} There is a unique solution $\mathbf{\Phi} (\mathcal{F}, \mathbf{R}^0)$ that satisfies all six axioms in the above, and is given by,
\begin{equation}
\mathbf{\Phi} (\mathcal{F}, \mathbf{R}^0) = \arg \max _{\mathbf{R} \in \mathcal{F}, \mathbf{R}\geq \mathbf{R}^0}\prod _{i =1}^2 (R_i - R^0_i)
\end{equation}
}
The NBS selects the unique allocation that maximizes the Nash product in (12) over all feasible individual rational allocations. Note that for any essential bargaining problem, the Nash point should always satisfy $R^*_i > R^0_i, \forall i$.

\section{Bargaining over the Two-User Gaussian MAC}
Before we move to the Gaussian IC, we first consider a Gaussian MAC in which two users send information to one common receiver. This also forms the foundation for the solution of the strong IC. The received signal is given by
\begin{equation}
Y = X_1 + X_2 + Z
\end{equation}
where $X_i$ is the input signal of user $i$ and $Z$ is Gaussian noise with zero mean and unit variance. Each user has an individual average input power constraint $P_i$. The capacity region $\mathcal{C}$ is the set of all rate pairs $(R_1, R_2)$ such that
\begin{eqnarray}
R_i \leq C(P_i), \: i \in \{1,2\}\\
R_1 + R_2 \leq \phi_0 = C(P_1 + P_2)
\end{eqnarray}
If the two users fully cooperate in codebook and rate selection, any point in $\mathcal{C}$ is achievable. When there is no coordination between users, in the worst case, one user's signal can be treated as noise in the decoding of the other user's signal, leading to rate $R_i^0 = C(\frac{P_i}{1+P_{3-i}})$  for user $i$. In \cite{references:Gajic08}, $R_i^0$ is also called user $i$'s ``safe rate''. If the two users are selfish but willing to coordinate for mutual benefits, they may bargain over $\mathcal{C}$ to obtain a fair operating point with $\mathbf{R}^0$ serving as a disagreement point. 

{\proposition There exists a unique NBS for the bargaining problem $(\mathcal{C}, \mathbf{R}^0)$, given by $\mathbf{R}^* = (R_1^0 + \frac{1}{\mu_1}, R_2^0 + \frac{1}{\mu_1})$ where $\mu_1 = \frac{2}{\phi_0-R_1^0-R_2^0}$.
}
\begin{proof}
Maximizing the Nash product in (12) with $\mathcal{F}$ replaced by $\mathcal{C}$ is equivalent to maximizing its logarithm.
Define $f(\mathbf{R}) = \ln(R_1 - R_1^0) + \ln(R_2 - R_2^0)$, then $f(\cdot): \mathcal{C}\cap \{\mathbf{R}|\mathbf{R}\geq \mathbf{R}^0\} \rightarrow \mathbb{R}^+$ is a strictly concave function of $\mathbf{R}$. Also note that the constraints in (12), (14) and (15) are linear in $R_1$ and $R_2$. So the first order Karush-Kuhn-Tucker conditions are necessary and sufficient for optimality \cite{references:Bertsekas}.
Let $L(\mathbf{R},\lambda,\mu)$ denote the Lagrangian where $\lambda_i \geq 0,\: i = 1,2$ and $\mu_1 \geq 0$ denote the Lagrange multipliers associated with the constraints, then we have
\begin{equation}
L(\mathbf{R},\lambda,\mu) = f(\mathbf{R}) + \sum_{i = 1}^2 \lambda_i(R_i - R_i^0)  + \mu_1 (\phi_0 - R_1 - R_2)
\end{equation}
The first-order necessary and sufficient conditions yield
\begin{equation}
1 + \left(\lambda_i - \mu_1\right)(R_i^* - R_i^0) = 0; \: i = 1,2
\end{equation}
and
\begin{eqnarray}
(R_i^* - R_i^0)\lambda_i = 0; \quad \lambda_i \geq 0; \: i = 1,2\\
(R_1^* + R_2^* - \phi_0)\mu_1 = 0; \quad \mu_1 \geq 0
\end{eqnarray}
Since $R_i^* > R_i^0$ must hold, we have $\lambda_i = 0$ for $i = 1,2$. Also the constraint $\phi_0 - R_1^* - R_2^* \geq 0$ has to be active, i.e.,
\begin{equation}
R_1^* + R_2^* = \phi_0
\end{equation}
Solving (17) and (20), we obtain $\mu_1 = \frac{2}{\phi_0-R_1^0-R_2^0}$ and $R_i^* = R_i^0 + \frac{1}{\mu_1},\: i = 1,2$ .
\end{proof}
In Fig. 1, the capacity region, the disagreement point and the NBS obtained using Proposition 1 are illustrated for $\text{SNR}_1 = 15$dB and $\text{SNR}_2 = 20$dB. Recall that the mixed strategy NE in \cite{references:Gajic08} has an average performance equal to the safe rates in $\mathbf{R}^0$. The NBS point which is the unique fair Pareto-optimal point in $\mathcal{C}$ is component-wise superior. This shows that bargaining can improve the rates for both of the selfish users in a MAC.
\begin{figure}
\centering
\includegraphics[width = 2.2in]{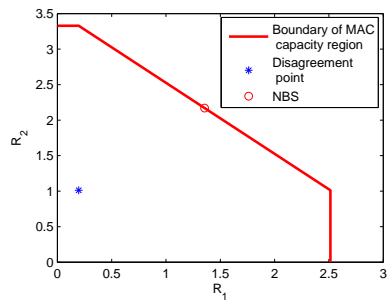}
\caption{The Nash point over the MAC when $\text{SNR}_1 = 15$dB, $\text{SNR}_2 = 20$dB}
\end{figure}

\section{Two-User Gaussian IC}
For the IC, the coordination between the two users is done in two phases. In phase 1, they negotiate for a simple H-K type scheme that has the potential to improve individual rates for both. The private message power factors $\alpha$, $\beta$ in the H-K scheme are jointly determined by both users and depend on their power constraints $P_1$, $P_2$ and channel parameters $a$ and $b$. If at least one user does not have the incentive to cooperate in the sense of Sec II-C, then negotiation breaks down; otherwise, they reach an agreement on the use of the H-K type scheme with the chosen power split. In phase 2, both users bargain for a fair rate pair in the bargaining set which is the achievable rate region of the H-K scheme they agreed on earlier. This problem can then be formulated as a two-user bargaining problem with the feasibility set $\mathcal{F}$ and disagreement point $\mathbf{R}^0$. Once a particular rate pair is determined as the solution, related codebook information is shared between the users so that one user's receiver can decode the other user's common message as required by the adopted H-K scheme in agreement. If negotiation breaks down, each receiver is not provided with the interfering user's codebook.

\subsection{Conditions for users to have incentives to cooperate}

In this subsection, we discuss the incentive conditions for both users to cooperate and how they jointly choose $\alpha$ and $\beta$ for different interference regimes. In the first phase, the two users search for a H-K scheme that could result in a rate region containing rate pairs component-wise better than $\mathbf{R}^0$. Intuitively, it would be best to have a scheme that could achieve the largest rate region that includes $\mathbf{R}^0$. While the full H-K achievable region \cite{references:Han81} needs to take into account all possible power splits and different time-sharing strategies, it is computationally infeasible. For tractability, we restrict the two users' choices to a simple H-K type scheme with fixed power split and no time-sharing. For the weak and mixed interference cases, we study incentive conditions for cooperation based on the near-optimal power split of \cite{references:Etkin08}. For the strong interference case, we set $\alpha = \beta = 0$, which is known to be optimal\cite{refereces:Sato81}.

\subsubsection{Strong Interference}
Suppose $a \geq 1$ and $b \geq 1$, and we choose optimal $\alpha = \beta = 0$. Treating interference as noise is suboptimal and $\mathbf{R}^0$ always lies inside $\mathcal{F}$. The bargaining problem $(\mathcal{F},\mathbf{R}^0)$ is essential and hence both users always have incentives to cooperate.

\subsubsection{Mixed Interference}
Without loss of generality, we assume $a < 1$ and $b \geq 1$. We use the near-optimal power splits $\alpha = 0$ and $\beta = \min(1/(aP_2),1)$\cite{references:Etkin08}. If $aP_2 \leq 1$, the interference from user 2 has a smaller effect on user 1 than the noise at user 1's receiver does. The scheme $\text{HK}(0, 1)$ will not improve user 1's rate and hence user 1 does not have an incentive to cooperate using this scheme. But if $aP_2>1$ and $\mathcal{F}\cap \{\mathbf{R}>\mathbf{R}^0\}$ is nonempty when $\alpha = 0$ and $\beta = 1/(a P_2)$, it is possible to improve both users' rates relative to those in $\mathbf{R}^0$.

Note that $aP_2> 1$ holds when $a > 1/P_2$. When $\text{SNR}_2$ is high, this condition is satisfied for most $a$'s. This implies that in the interference limited regimes, it is very likely that both users would have incentives to cooperate. The case for $a \geq 1$ and $b<1$ can be analyzed similarly.

\subsubsection{Weak Interference}
Suppose $a <1$ and $b<1$. We use the power splits $\alpha = \min(1/(bP_1),1)$ and $\beta = \min(1/(aP_2),1)$\cite{references:Etkin08}. Similar to the mixed case, only if $aP_2 > 1$, $bP_1 > 1$ and $\mathcal{F}\cap \{\mathbf{R}>\mathbf{R}^0\}$ is nonempty when $\alpha = 1/(bP_1)$ and $\beta = 1/(aP_2)$, both users' rates can be improved compared with those in $\mathbf{R}^0$. 

Note that as in the mixed case, when both $\text{SNR}$'s are high, the conditions $aP_2 > 1$ and $bP_1 > 1$ are satisfied for most channel gains in the range and it only remains to check whether $\mathcal{F}\cap \{\mathbf{R}>\mathbf{R}^0\}$ is nonempty.

\begin{figure*}[ht]
\centering
\subfigure[Strong interference $a = 3$, $b = 5$]{
\includegraphics[width = 2in]{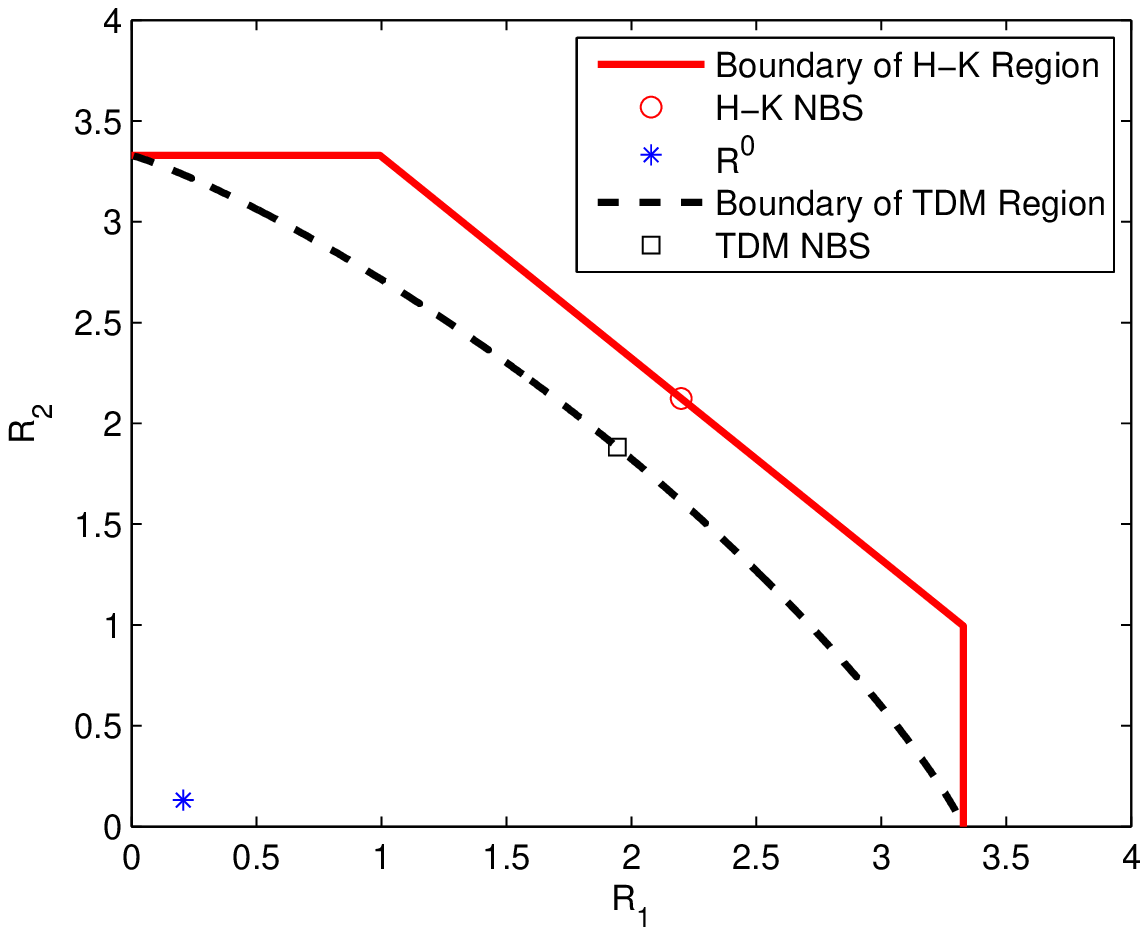}
\label{fig:subfig1}
}
\subfigure[Mixed interference $a = 0.1$, $b = 3$]{
\includegraphics[width = 2in]{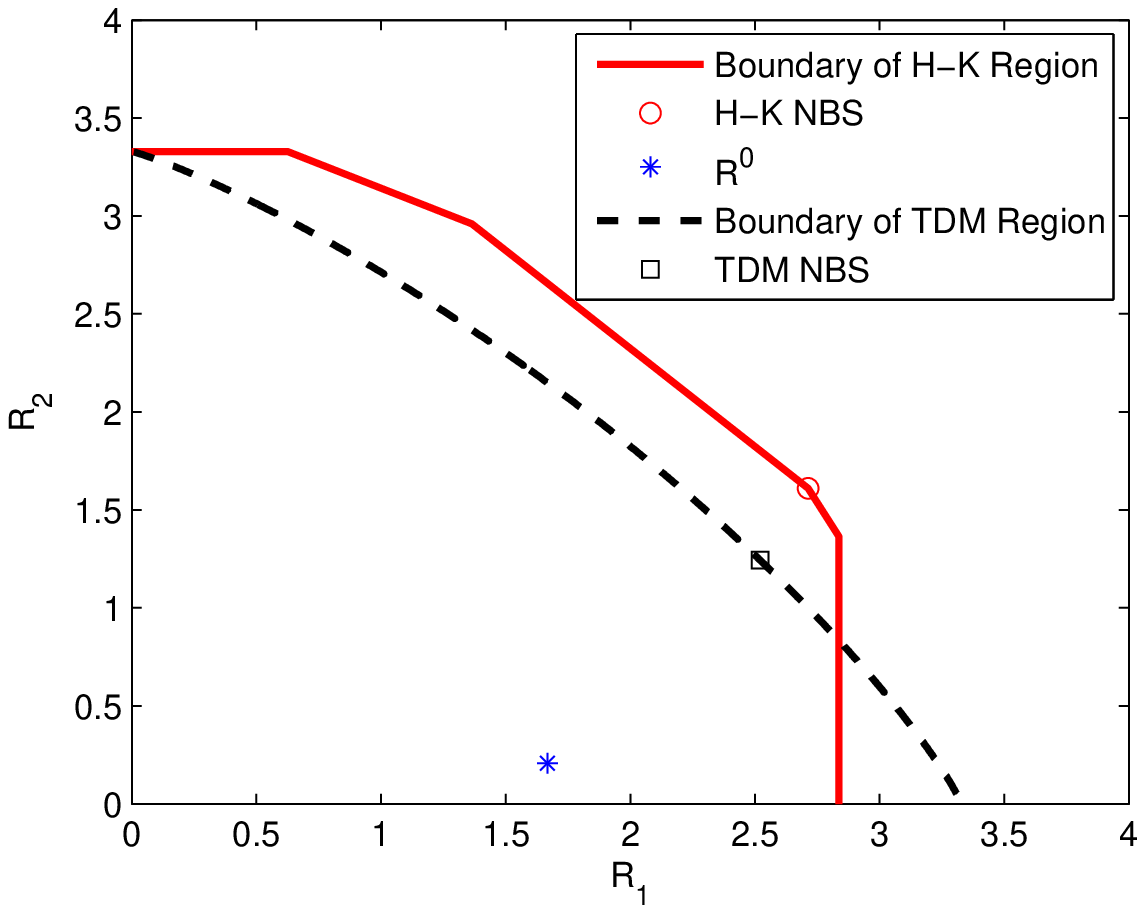}
\label{fig:subfig3}
}
\subfigure[Weak interference $a = 0.2$, $b = 0.5$]{
\includegraphics[width = 2in]{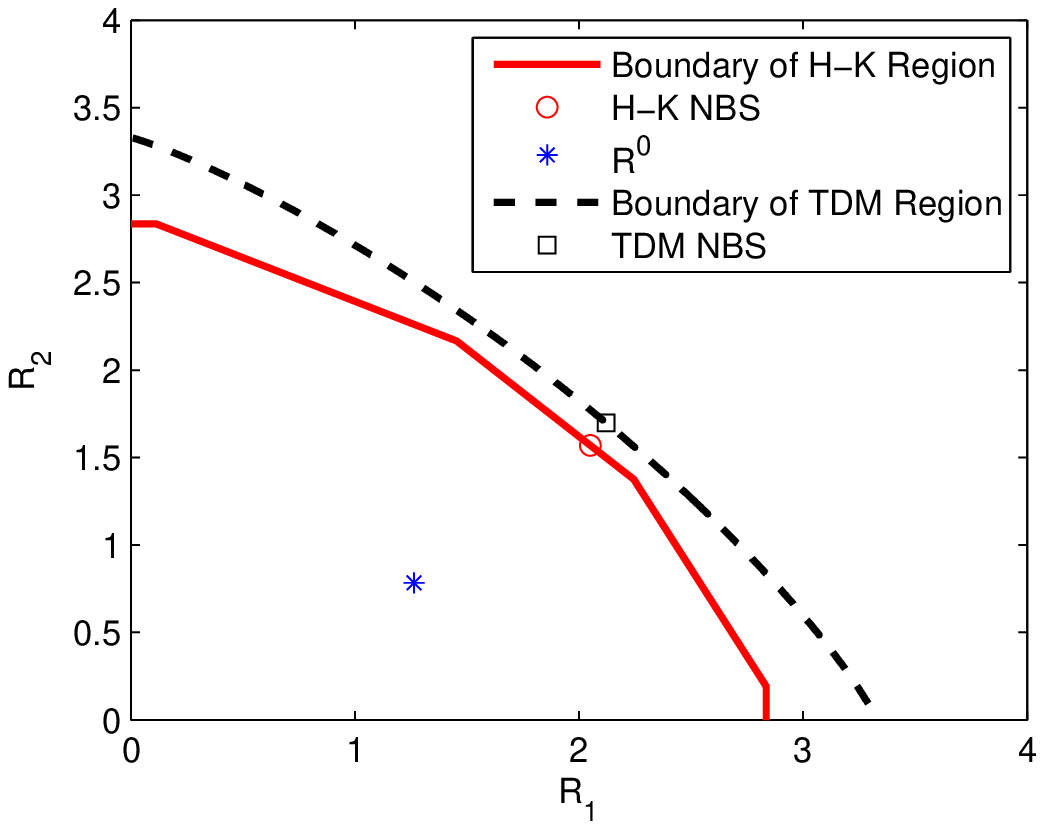}
\label{fig:subfig2}
}
\label{fig:subfigureExample}
\caption{The H-K NBS of the Gaussian IC in different interference regimes when $\text{SNR}_1 = \text{SNR}_2 = 20$dB.}
\end{figure*}

\subsection{Computing the Nash Bargaining Solution}
After the users agree on an H-K scheme, in phase 2, the NBS over the corresponding rate region $\mathcal{F}$ is employed as the operating point. We concentrate on the case when $\mathbf{R}^0<\mathbf{R}^1\:\text{and} \:\mathbf{A}\mathbf{R}^0<\mathbf{B}$ so that $\mathcal{F}\cap \{\mathbf{R}>\mathbf{R}^0\}$ is nonempty. From Section II, we know that, the NBS exists for $(\mathcal{F},\mathbf{R}^0)$ and is unique. It can be computed by optimizing (12).

{\proposition Assuming that $\mathbf{R}^0<\mathbf{R}^1\:\text{and} \:\mathbf{A}\mathbf{R}^0<\mathbf{B}$, there exists a unique NBS $\mathbf{R}^*$ for the bargaining problem $(\mathcal{F},\mathbf{R}^0)$, which is characterized as follows:
\begin{equation}
R_i^* = \min \left\{ R_i^1; R_i^0 + \frac{1}{\sum_{j = 1}^3 \mu_j A_{ji}}\right\}, \quad i\in\{1,2\}
\end{equation}
where $\mu_j \geq 0$, $j \in \{1,2,3\}$ is chosen to satisfy
\begin{equation*}
 (\mathbf{AR}^*-\mathbf{B})_j \mu_j = 0, \quad \mathbf{A}\mathbf{R}^* \leq \mathbf{B}
\end{equation*}
}
\vspace{-0.2cm}
\begin{proof}
As in the proof of Proposition 1, we use the Lagrange multiplier method and Karush-Kuhn-Tucker optimality conditions. Due to limited space, the proof is omitted.
\end{proof}

\section{Illustration of Results}
The achievable rate region of the H-K scheme and the H-K NBS are plotted for different values of channel coefficients in Fig. 2. For comparison, we also include the TDM region and the TDM NBS. The TDM region is given by $\mathcal{R}_{\text{TDM}}= \{\mathbf{R}|\mathbf{R} = (\rho_1 C(\frac{P_1}{\rho_1})\;\rho_2 C(\frac{P_2}{\rho_2}))^t,\:\rho_i\geq 0, \forall i,\:\rho_1+\rho_2\leq 1\}$ and the TDM NBS is computed by optimizing (12) with $\mathcal{F}$ replaced by $\mathcal{R}_{\text{TDM}}$. The NBS based on TDM was also investigated in \cite{references:Leshem08} using the unique competitive solution studied there as the disagreement point. Since interference limited regimes are more of interest here, in these plots, we assume both SNR's are high, i.e, $\text{SNR}_1 = \text{SNR}_2 = 20$dB. In Fig. 2(a), both interfering links are strong, hence $\text{HK}(0,0)$ is employed. The H-K NBS strictly dominates the TDM one. Fig. 2(b) shows an example for mixed interference case when $a = 0.1$ and $b = 3$. Since $aP_2 = 10> 1$, $\text{HK}(0,0.1)$ is employed. In this example, although TDM results in some rate pairs that are outside the H-K rate region, the H-K NBS remains component-wise better\footnote{Note this may not necessarily hold for all $\text{SNR}$'s and the channel gains in the range. In other words, for some other parameters, it is possible that one user gets a higher rate in the TDM NBS than in the H-K NBS.} than the TDM one. The weak interference case when $a = 0.2$ and $b = 0.5$ is plotted in Fig. 2(c). Given these parameters, we have $aP_2 = 20>1$ and $bP_1 = 50>1$, therefore $\text{HK}(0.02,0.05)$ is used. The H-K NBS, though still much better than $\mathbf{R}^0$, is slightly worse than the TDM one. This is because the TDM rate region contains the H-K rate region due to the suboptimality of the simple H-K scheme in the weak regime. Note that we do not employ time sharing in the chosen H-K scheme. Finally, recall that while the TDM rate region does not depend on $a$ and $b$, since $\mathbf{R}^0$ does, the TDM NBS depends on $a$ and $b$ as well.

We compute the H-K NBS for different ranges of the channel gains in Fig. 3. We assume $\text{SNR}_1 = \text{SNR}_2 = 20$dB, $a = 1.5$ and $b$ varies from 0 to 3. For all $b$'s, both users' rates in the NBS $\mathbf{R}^*$ are higher than those in $\mathbf{R}^0$. The improvement of each user's rate in $\mathbf{R}^*$ over the one in $\mathbf{R}^0$ increases as $b$ grows. When $b< a$, user 1's rate in the NBS is less than user 2's; however, as $b$ grows beyond $a$, user 1's rate in the NBS surpasses user 2's, which is due to the fairness property of the NBS. Alternatively we say a strong interfering link can give user 1 an advantage in bargaining. In Fig. 4, we plot the sum rates for H-K NBS and TDM NBS under the same setting as in Fig. 3. For comparison, the maximum sum rate of the H-K scheme with the chosen power split is also given. The H-K NBS performs better in terms of sum rates than the TDM NBS for all $b$'s except when $b$ is around 1, where the performances of the two schemes are similar. Moreover, the H-K NBS rate pair can achieve the maximum sum rate of the H-K scheme used for almost all $b$'s except when $b$ is very small ($\leq 0.05$), the sum rate of the H-K NBS is relatively lower. This demonstrates that the H-K NBS not only provides a fair operating point but also maintains a good overall performance.

\begin{figure}
\centering
\includegraphics[width = 2.5in]{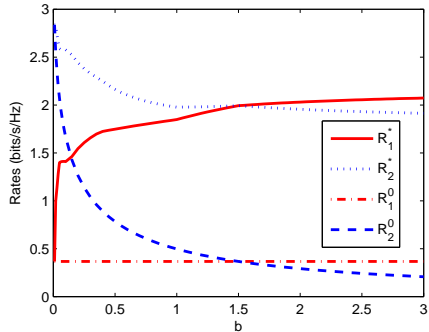}
\caption{Rates in the NBS $\mathbf{R}^*$ and disagreement point $\mathbf{R}^0$ when $\text{SNR}_1 = \text{SNR}_2 = 20$dB and $a = 1.5$.}
\end{figure}

\begin{figure}
\centering
\includegraphics[width = 2.5in]{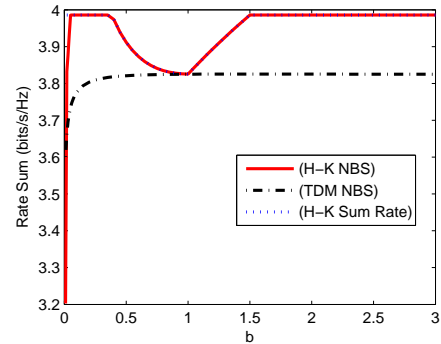}
\caption{Sum rates in the H-K NBS when $\text{SNR}_1 = \text{SNR}_2 = 20$dB and $a = 1.5$.}
\end{figure}
\section{Conclusions}
In this paper, we investigated the two-user Gaussian IC, under the assumption that the two users are selfish and interested in cooperation only when they have incentives to do so. We proposed a two-phase mechanism for the two users to coordinate, which consists of choosing a simple H-K type scheme with Gaussian codebooks and fixed power split in phase 1 and bargaining over the achievable rate region to obtain a fair operating point in phase 2. We show that the proposed mechanism can gain substantial rate improvements for both users compared with the uncoordinated case. The obtained operating point is also strongly efficient in the sense that it can achieve the maximum sum rate of the adopted simple H-K type scheme in most cases.

\bibliographystyle{IEEEtran}
\bibliography{IEEEabrv,references}
\end{document}